\documentclass[12pt]{article}

\usepackage{amsmath,amssymb,amsfonts}

\usepackage{times}
\usepackage{bm,mathrsfs,dsfont,color}
\usepackage{natbib}
\usepackage{graphicx}
\usepackage{subfigure}
\usepackage[font=small]{caption}

\usepackage{graphicx}
\usepackage{amsmath}
\usepackage{amsfonts}
\usepackage{enumitem}
\usepackage{comment}
\usepackage{xurl}

\usepackage{bbm}
\usepackage{threeparttable}

\usepackage{authblk}

\allowdisplaybreaks

\usepackage{multirow}

\graphicspath{{img/}}

\usepackage[plain,noend]{algorithm2e}

\makeatletter
\renewcommand{\algocf@captiontext}[2]{#1\algocf@typo. \AlCapFnt{}#2} 
\def\@algocf@capt@plain{top}
\renewcommand{\algocf@makecaption}[2]{%
	\addtolength{\hsize}{\algomargin}%
	\sbox\@tempboxa{\algocf@captiontext{#1}{#2}}%
	\ifdim\wd\@tempboxa >\hsize
	\hskip .5\algomargin%
	\parbox[t]{\hsize}{\algocf@captiontext{#1}{#2}}
	\else%
	\global\@minipagefalse%
	\hbox to\hsize{\box\@tempboxa}
	\fi%
	\addtolength{\hsize}{-\algomargin}%
}
\makeatother

\begin{document}

	\makeatletter
	\renewcommand\section{\@startsection {section}{1}{\z@}%
		{-3.5ex \@plus -1ex \@minus -.2ex}%
		{2.3ex \@plus.2ex}%
		{\centering\normalfont\large\scshape}}
	
	\renewcommand\subsection{\@startsection {subsection}{1}{\z@}%
		{-3.5ex \@plus -1ex \@minus -.2ex}%
		{2.3ex \@plus.2ex}%
		{\centering \normalfont \scshape}}                                   
	\makeatother
	
	\title{Structured factorization for single-cell gene expression data}
	
	\author[1]{
		Antonio Canale\thanks{\tt antonio.canale@unipd.it }}
	\author[1]{Luisa Galtarossa}
	\author[1]{Davide Risso} 
	\author[2]{Lorenzo Schiavon}
	\author[1]{Giovanni Toto}

	\affil[1]{Department of Statistical Sciences, University of Padova, Padova, Italy}
	\affil[2]{Department of Economics, Ca' Foscari University of Venice, Italy}
	
	\date{}
	
	\maketitle
	
	\abstract{
Single-cell gene expression data are often characterized  by large  matrices, where the number of cells may be lower than the number of genes of interest. Factorization models have emerged as powerful tools to condense the available information through a sparse decomposition into lower rank matrices. In this work, we adapt and implement a recent Bayesian class of generalized factor models to count data and, specifically, to model the covariance between genes. 
The developed methodology also allows one to include exogenous information within the prior, such that recognition of covariance structures between genes is favoured. In this work, we use biological pathways as external information to induce sparsity patterns within the loadings matrix. 
This approach facilitates the interpretation of loadings columns and the corresponding latent factors, which can be regarded as unobserved cell covariates. We demonstrate the effectiveness of our model on single-cell RNA sequencing data obtained from lung adenocarcinoma cell lines, revealing promising insights into the role of pathways in characterizing gene relationships and extracting valuable information about unobserved cell traits. }
	
	{\center \textbf{Keywords: }}
Count data; Factor analysis; Pathways; Shrinkage prior; Rounded continuous data;

\newpage 
\section{Introduction}
\label{s:intro}

\subsection{Single-cell RNA sequencing data}
Single-cell RNA sequencing (scRNA-seq) has become a widely used tool to characterize gene expression of thousands of cells at transcriptome-wide resolution. By sequencing RNA molecules from individual cells, scRNA-seq provides a count-based measure of relative gene expression.
Compared to previous ``bulk'' technologies, single-cell sequencing unlocks the possibility to analyze rare cell types, to discover new cell types, and to study the heterogeneity of  gene expression in cell populations of interest \citep{wagner2016revealing}. This is particularly relevant in cancer, as tumours interact with the surrounding tissues, known as the tumour microenvironment; this interaction is associated with prognosis, response to treatment, and survival \citep{wu2017tumor}. Studying tumour samples at single-cell resolution allows for the discovery of cell sub-populations that potentially respond differently to treatment \citep{xue2020rapid} or that are differentially abundant across patients or disease stages \citep{becker2022single}, making it a promising tool for personalized medicine.

For each cell, scRNA-seq data consist of counts that represent the expression of each gene in that cell.
In a typical experiment, in addition to the cells by genes expression matrix, several supporting variables are collected for each of the analyzed cells and for each of the measured genes; we name ``covariates'' the former and ``meta-covariates'' the latter. 

We denote the matrix of gene expression as $y$; such matrix, of dimension $n \times p$, contains the counts for $p$ genes measured on $n$ cells. 
The matrix containing the covariates, $x$, is a $n \times d$ matrix where $d$ indicates the number of covariates for each cell. The covariates are cell-specific features, typically containing quality control information, such as the number of mapped or aligned reads and the total counts, as well as phenotypic information, such as the 
tissue or donor.
In addition, the matrix containing the meta-covariates is indicated with $w$; it has dimensions $p \times q$ and contains the $q$ meta-covariates for each gene, i.e. gene-specific features containing technical, e.g., gene length or GC-content, and biological, e.g., pathway membership, information. 

Gene expression data at single-cell resolution are highly informative, allowing researchers to characterize the cells at the finest level, and their application to cancer research, immunology, and developmental biology have already led to novel insights. However, scRNA-seq data are challenging: they are high-dimensional count data, characterized by high variance and abundance of zeros. Hence, exploratory models are needed to facilitate summary, visualization and clustering of cells, to identify novel biological hypotheses to be tested with targeted experiments.

\subsection{High dimensional count data challenges}

A default strategy for modelling RNA-seq count data consists in using standard parametric distributions such as the Poisson \citep{marioni2008}
or the negative binomial \citep{anders2010,robinson2008}.
Even if simple in terms of computation and interpretation, such standard models have some limitations. For instance, even the negative binomial may be unable to capture the zero inflation and multimodality of gene-wise distributions often observed in scRNA-seq \citep{jiang2022}.

A different -- unfortunately still common -- approach forgets the count nature of the data and treats them as continuous. A common practice, often used also in different applications in which count data are observed, consists of log- or square-root-transforming the observed counts, subsequently applying methods designed for continuous or Gaussian data. However, transformations to Gaussianity are ineffective for small counts \citep{warton2018}, while log-transformations introduce difficulties in the presence of zeros \citep{hara2010}. 
This practice has been strongly criticized in our motivating context of scRNA-seq data \citep[e.g.,][]{townes2019}. More broadly, these approaches are not well-defined for count data: the data-generating process for a continuously-transformed Gaussian model cannot produce counts, which immediately amplifies model misspecification, limits interpretability, and undermines the reliability of inference and predictive distributions.

To overcome these limitations and challenges, we introduce a flexible Bayesian framework specifically designed to handle complex count-valued data. Our proposed framework incorporates a continuous latent variable representation, similar to the approaches employed by  \citet{canale2011} and \citet{kowal}. By adopting this specification, we are able to effectively capture the various characteristics associated with high-dimensional count probability mass functions.
In addition, to account for the intricate dependence structures present in the multivariate count vector, we follow the common practice that leverages factorization models. This approach allows us to express the high-dimensional covariance matrix as a combination of a limited number of rank-one matrices. Recently, \citet{schiavon2022} introduced a general class of infinite factorization models capable of handling continuous, binary, and count data. Notably, this class of models promotes sparsity in the matrix of factor loadings by effectively incorporating information from covariate and metacovariate vectors.

In the next Section, we provide a detailed description of our proposed approach, which we refer to as  \textsc{cosin}  (COunt data Structured INfinite factorization). Subsequently, in Section \ref{s:application}, we apply this model to a motivating example involving lung adenocarcinoma scRNA-Seq data. To assess the validity and generality of our approach, as well as to compare its performance against state-of-the-art scRNA-seq methods, we present a comprehensive simulation experiment in Section \ref{s:simulations}. Finally, in Section \ref{s:discuss}, we engage in a thorough discussion of the proposed approach, its generalization and extensions, and its possible applications beyond scRNA-seq studies.

\section{Model and prior specification}
\label{s:model}

For each cell $i=1, \dots,n$, scRNA-seq data can be treated as a $p$-dimensional vector of integer valued random variables $y_i \in \mathbb{N}^p$ where $\mathbb{N}$ is the set of natural numbers. Along with the $n\times p$ data matrix $y$, additional external information for each cell and each gene are also available. Let $x_i$ be the $d$-dimensional vector of cell-specific covariates and $w_j$ for $j=1, \dots, p$ be the $q$-dimensional vector of gene-specific meta-covariates with  $w_j^\top = (w_{Tj}^\top, w_{Bj}^\top)$ where $w_{Tj}$ is the $q_T$ dimensional subvector of technical meta-covariates and $w_{Bj}$ is the $q_B$ dimensional subvector of biological meta-covariates and $q = q_T + q_B$.

\subsection{Model specification}

Following \citet{kowal} we introduce a continuous-valued latent matrix $z$ related to  the observed count-valued  matrix $y$ via a simultaneous transformation and rounding operator ${\cal S}\!: \mathbb{R} \to \mathbb{N}$ with ${\cal S}(\cdot) = {\cal H}({\cal G}(\cdot))$. Specifically, the rounding operator is such that ${\cal H}(t) = \ell$ if $t \in \mathcal{A}_\ell$ and $\{\mathcal{A}_\ell\}_{\ell=0}^\infty$ is a known partition of $\mathbb{R}$. Here, we adopt  the floor function defined by $\mathcal{A}_\ell = [\ell, \ell+1)$. As discussed in \citet{kowal}, rounding alone is suboptimal, particularly when the original data are counts. The popularity of log-linear models for count data thus suggests to specify the transformation operator $\cal G$ as the exponential transformation. Thus, the single entry $y_{ij}$ of $y$ is linked to a latent $z_{ij}$ via the  operator $\cal S$ and specifically $y_{ij} = \ell$ if $\exp\{z_{ij}\} \in [\ell, \ell+1)$. 

The latent variables $z_{ij}$ are modelled via 
\begin{equation}\label{eq:simpleGP}
z_{ij} = x_i^T \beta_j + \epsilon_{ij},
\end{equation}
where $x_i^T \beta_j$ is the conditional expectation of $z_{ij}$ and  $\epsilon_{ij}$ is a zero-mean  Gaussian error term. Note that we are assuming a linear relation between the mean of the latent variables and the set of cell-specific covariates $x_i$ and that this relation is changing with $j$, i.e. we assume that the same set of covariates may impact differently the different columns of the matrix $z$. 

The Gaussian error term captures all the residual variability not modeled by the linear predictor in \eqref{eq:simpleGP}. Consistently with this, we exploit a factor analytic representation that allows us to express $\epsilon_{i}$ as the linear combinations of latent $k$-dimensional factors $\eta_i$. More formally we let
\begin{equation}\label{eq:simpleFactor}
\epsilon_{ij} = \sum_{h= 1}^{k} \lambda_{jh} \eta_{ih} + \varepsilon_{ij},
\end{equation}
where  $\lambda_{jh}$ is an element of the $p \times k$ factor loadings matrix $\Lambda$ and $\eta_{ih}$ is an element of the $h$-th latent factor $\eta_{\cdot h}$, with $h=1,\ldots,k$. The vectors $\varepsilon_i$ represent the remaining noise and are iid according to a $p$-variate Gaussian distribution $N(0, \Sigma)$, with diagonal covariance matrix $\Sigma$.
In matrix notation, the error matrix $\epsilon$ is equal to a sum of $n \times p$ rank-one matrices identified by the vector product $\eta_{\cdot h} \lambda_{\cdot h}^\top$. We refer to such matrices as rank-one additive contributions $C_h$.
Notably, if the number $k$ of these contributions is $k \leq p$, the factor representation leads to a parsimonious model.

\subsection{Prior specification}
Following a Bayesian approach we elicit suitable prior distributions for the model parameters.
%
The conditional expectation of $z_{ij}$ is modelled through a linear combination of a vector of cell-specific covariates $x_{i}$ weighted by a vector of regression parameter $\beta_j$, which differs over the genes $j=1,\ldots, p$. The availability of gene-specific prior information $w_j$ allows one to model the regression parameters accordingly. Indeed, one may expect that the expression of a gene in a certain cell depends on the cell traits $x_i$, but with such relation varying according to the gene characteristics $w_j$. It is well known, for example, that the gene length and its sequence composition (e.g., the proportion of guanine and cytosine nucleotides, known as GC-content) influence gene expression quantification, potentially in sample-specific ways \citep{risso2011gc,love2016modeling}. Hence, we specify
\begin{equation*}
\beta_j \sim N_{d}( \Gamma_T w_{jT}, \sigma_\beta^2 I_d)
\end{equation*}
where $\Gamma_T$ is a $d \times q_T$ coefficient matrix that model how the technical characteristics $w_T$ of the genes impact the cell quality control parameter.
In multivariate regression, such hierarchical structure on the mean process is common when additional information on the column entities is available \citep[see, e.g.][for ecological applications]{ovaskainen2020}. For instance,
one may expect that the impact of the number of mapped reads 
on the expression of gene $j$ varies according to the gene's technical traits.
We set the prior of $\Gamma_T$ entries as independent standard Gaussian random variables.

Inspired by such structure on the mean, we exploit the structured increasing shrinkage prior introduced by \citet{schiavon2022} to induce a gene-specific effect also on the loadings $\Lambda$, which model the impact of the latent cell traits $\eta_{\cdot h}\,(h=1,\ldots,k)$.
Consistently with this, the variance of each loading element is decomposed through the product of a factor-specific scale $\theta_h$ and a local scale $\phi_{jh}$ leading to the following  hierarchical prior
\begin{align*}
\lambda_{jh} \sim N(0, \theta_{h} \phi_{jh}),\quad \theta_h = \vartheta_h \rho_h,\\
\vartheta_h^{-1} \sim \text{Ga}(a_\theta, b_\theta), \quad \rho_h\ \sim \text{Ber}(1-\pi_h),
\end{align*}
where $\text{Ga}(a_\theta, b_\theta)$ indicates a gamma distribution with mean $a_\theta/b_\theta$ and variance $a_\theta/b_\theta^2$ and $\text{Ber}(1-\pi_h)$ is a Bernoulli distribution with mean $1-\pi_h$.

In such construction, $\pi_h$ is the probability of factor $h$ being shrunk to zero and is defined according to the stick-breaking construction 
\begin{equation*}
\pi_h = \sum_{l=1}^h u_l, \quad u_l = v_l \prod_{m=1}^{l-1} (1-v_m), \quad v_m \sim \text{Be}(1,\alpha),
\end{equation*}
where Be$(a,b)$ indicates the beta distribution with mean $a/(a+b)$. Under this cumulative construction, $\pi_{h+1} > \pi_h$ for any $h>0$ and $\lim_{h \rightarrow \infty} \pi_h = 1$ almost surely. The probability of being shrunk is increasing over the the index $h=1,\ldots,H$ allowing for an infinite factorization model \citep{bhattacharya2011} when $k$ is set equal to $+\infty$, which can be approximated by a truncated version of the same model.
\citet{legramanti2020}, which firstly introduced the cumulative stick-breaking construction to define a class of infinite factor models, note that the prior expected number of non
shrunk columns of $\Lambda$ is $E(\sum_{k=1}^\infty \rho_k)=\alpha$, suggesting setting $\alpha$ equal to the expected number of active latent factors. 

The scale $\phi_{jh}$ has a Bernoulli prior distribution and regulates the local shrinkage of the loadings. We model the local behaviour according to the mean equation
\begin{equation*}
E(\phi_{jh}) = c_p \text{logit}^{-1}(w_{jB}^\top \gamma_{hB}), \quad \gamma_{hB}\sim N(0, \sigma^2_\gamma I_q),
\end{equation*}
where logit$^{-1}(x)=e^x/(1+e^x)$, $c_p\in(0,1)$ is a possible offset, and $\gamma_{hB}$ is the $h$th column vector of a $q_B  \times k$ matrix $\Gamma_B$ with independent standard Gaussian prior. 
The vector $w_{jB}$ represents the realization of $q_B$ available gene-specific meta-covariates that we think could influence the effect of the latent unobserved covariates $\eta_k\,(k=1,\ldots,d)$, correspondingly to their role in the specification of the covariate effects $\beta$. 
Coefficients of the unobserved covariates are shrunk jointly in
similar genes, i.e. with similar meta-covariates.  
In particular, we consider as meta-covariates the binary 
vector $w_{jB}$ that indicates the biological pathways including gene $j$. 
We use pathways meta-covariates here, as we expect the factor loadings to be influenced by the biological processes that genes contribute to. In other words, we expect that genes that interact in a given biological process act in a coordinated way in defining the factors inferred by our model. We use pathway membership as a proxy for biological process, as usually done in bioinformatics \citep{khatri2012ten}.


\subsection{Posterior computation and point estimation}
\label{sec:posteriorcomputation}
Bayesian inference uses the posterior distribution of model parameters, which is approximated through Markov chain Monte Carlo (MCMC) sampling. Following common practice in infinite factor models \citep{bhattacharya2011,legramanti2020,schiavon2022}, we use an adaptive Gibbs algorithm to infer the number of active factors while drawing from the posterior distribution. To ensure convergence of the Markov chain, as stated in Theorem 5 of \citet{roberts2007}, the value of the number of factors is adapted at certain iterations with exponentially decreasing probability.

At the adaptive iterations, active factors are identified as those characterized by non zero loadings column -- i.e. $\rho_h=1$ -- and the redundant factors are discarded.
Given the number of factors $k$ at a certain iteration, model parameters are drawn from the corresponding posterior full conditional distributions. The detailed
steps of the adaptive Gibbs sampler are reported in  Appendix A in the supporting information.

In Bayesian analysis, point-wise estimates are usually obtained by approximating the parameters' posterior expectations via Monte Carlo averages over the samples drawn during the MCMC.
However, it is well-known in the Bayesian factor model literature that the sample average cannot informatively summarize the posterior distribution of $\Lambda$ and $\eta$, due to their non-identifiability.
In fact, both $\Lambda$ and $\eta$ are only identifiable up to an arbitrary rotation $P$ with $PP^\top = I_k$, causing sampling of such parameters from possibly different rotational alignments in different Gibbs iterations.
Given the sign symmetry of possible rotations, Monte Carlo averages would result in poorly meaningful point-wise estimates around zero.
On the other hand, the non-identifiable possible rotations of the rank-one contributions $C_h = \eta_{\cdot h} \lambda_{\cdot h}^\top$ are limited to the class of permutations of the indices $h=1,2,\ldots$. Then, focusing on rank-one contributions, identifiability is achieved by following the steps below.
\begin{enumerate}[label= \textit{\roman*.}, leftmargin=2\parindent]
	\item Order the contributions $C_1^{(T)},\ldots,C_k^{(T)}$ sampled at the last iteration $T$ of the Gibbs algorithm 
	decreasingly with respect to the Frobenius norm.
	\item Use the re-ordered contributions of the last iteration $C_{1^*}^{(T)},\ldots,C_{k^*}^{(T)}$ as a reference.
	\item For each Gibbs iteration $t<T$, re-order the contributions as follows.
	For $h^*=1,\ldots,k$, the contribution $C_{h^*}$ corresponds to the non ordered Contribution $C_h$ with  index 
	\begin{equation*}
	h = \text{argmin}_{l \in \mathbb{H}_h} ||C_{h^*}^{(T)} -  C_{l}^{(t)}||_F,
	\end{equation*}
	where $\mathbb{H}_h$ is the set of $k-h^*+1$ indices of non re-ordered contributions and $||A||_F$ denotes the Frobenius norm of matrix $A$.
\end{enumerate}
To obtain point-wise estimates, we compute, for each $h=1,\ldots,k$, the sample mean $\bar{C}_h=(\sum_{t=1}^T C_{h^*}^{(t)})/T$ over the re-ordered contributions.
To investigate the behaviour of the factors scores $\eta$, we select a representative iteration of the Gibbs sampler by
following the procedure described in \citet{schiavon2022}. For alternative post-processing algorithms to align the samples of $\Lambda$ or $\eta$ we refer the reader to \citet{mcparland2014}, \citet{assmann2016}, and \citet{roy2019}.

\section{Lung adenocarcinoma scRNA-seq data}
\label{s:application}

We analyze a subset of the study of three lung adenocarcinoma cell lines measured by scRNA-seq by \cite{tian2019}. The original data are high quality in terms of exon mapping and unique transcript counts per cell and have already undergone pre-processing and quality control by the original authors using the \texttt{scPipe} workflow \citep{scpipe}. 
Cells with high percentages of mitochondrial genes and less than 1000 detected genes are excluded and only genes belonging to the 20 largest pathways are retained. The resulting data consist of a gene expression matrix of $n=199$ cells and $p=949$ genes. Along with this gene expression matrix, quality control information about the cells $x$ and the genes $w_T$ are also available. The matrix $x$ includes technical features, such as the number of unaligned reads, the number of reads mapped to exons or introns, the number of expressed genes, and the total number of counts, leading to $d=9$ covariates.  
In addition, cell line information is also available, indicating for any cell if it belongs to cancer lines H1975, H2228, or HCC827. We chose not to use such information as a covariate in the model, to mimic a typical situation in scRNA-seq, in which the identity of the cells is not known in advance. Hence, it represents a useful benchmark to assess the capacity of the model to reconstruct unobserved covariates in high-dimensional settings.
Technical meta-covariate matrix $w_T$ includes the length and the GC-content of each gene.
Consistently with the motivations previously mentioned, for each gene $j$, we also define a biological meta-covariate binary vector $w_{jB}$ with $m$ entry equal to $1$ if the gene $j$ belongs to the $m$ biological pathway. The list of $q_B=20$ pathways considered is reported in  Table A2 of the Appendix.

We  apply  \textsc{cosin}  with latent continuous Gaussian variable $z$ specified as in \eqref{eq:simpleGP}--\eqref{eq:simpleFactor}. Parameter prior distributions follow the hierarchical structure discussed in Section \ref{s:model}.
Given the high dimensionality of the data set, we may expect a sufficiently large number of latent factors, hence we set $\alpha=10$.
To favour variable selection, we shrink covariate coefficients setting $\sigma_\beta^2=1/3$. Additional details and MCMC settings are reported in the supporting information.

\begin{figure}
	\centerline{
		\includegraphics[width=.9\textwidth]{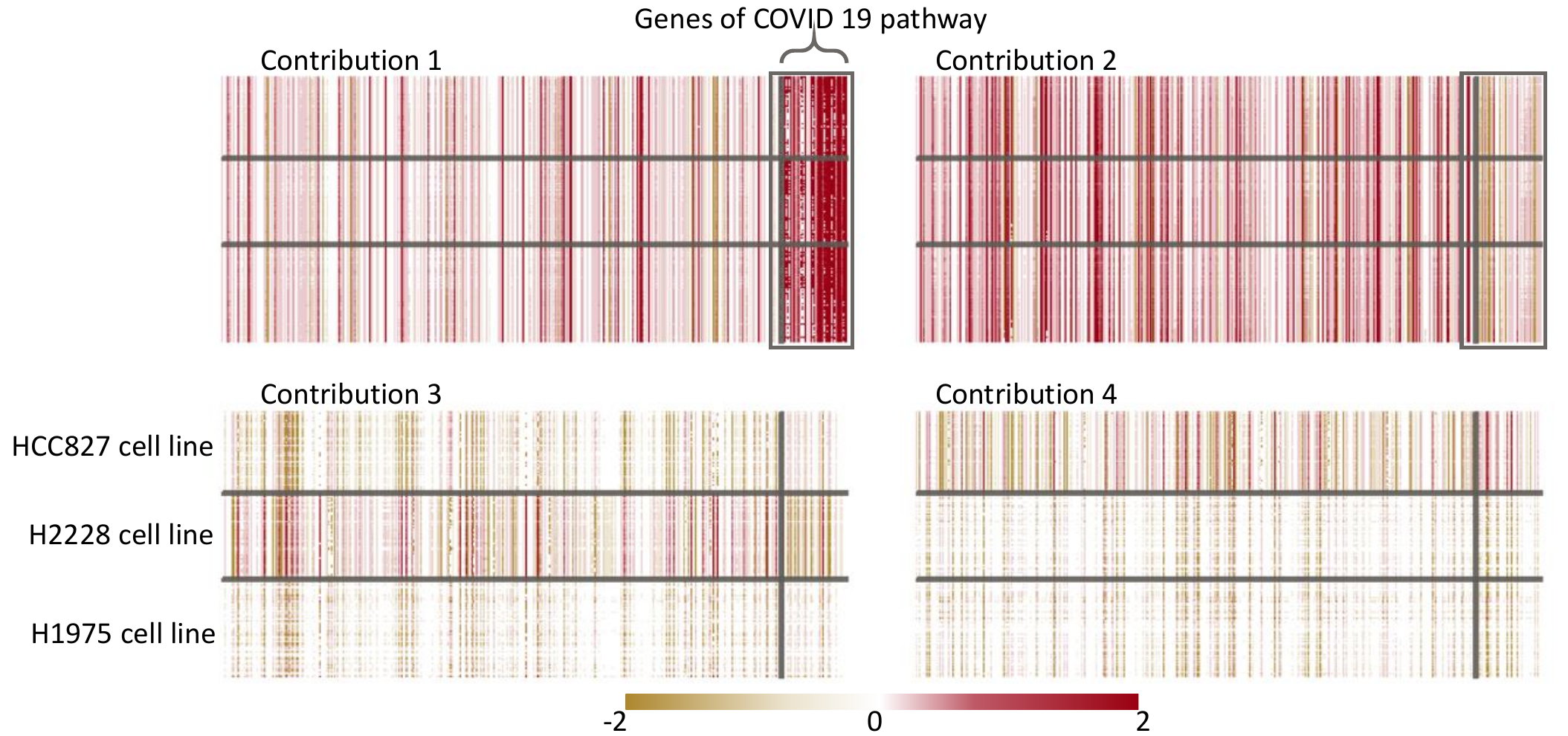}}
	\caption{First four rank-one contribution matrices. Genes are ordered according to the COVID-19 disease biological pathway: the genes within the grey square on the right of each contribution matrix belong to the COVID-19 pathway. Rows are re-arranged according to cell lines.}
	\label{fig:cont-covid}
\end{figure}

First, we discuss the results obtained for the mean of the process and the cell covariate effects. 
Summaries of the matrix  $\beta$ are reported in Table A3 in the Appendix.
The last column illustrates that only a small proportion of $\beta$'s for any covariate has $0.9$-level posterior credible intervals not including zero. This fact suggests a low importance of the control quality cell features, confirming that the  data are of high quality. 
As one may expect, the total number of genes detected per cell seems positively associated to the gene expression, but with just few genes where such effect is relevant. 
The variables describing the number of reads mapped to introns and to mitochondrial genes are those relevantly impacting the highest number of genes. 
The direction of such impacts varies on different genes, with variability partially explained by the technical characteristics of the genes, i.e. the length and the GC-content of the genes.

 Figure A5, reported in the Appendix, displays the posterior distributions of the $\Gamma_T$ coefficients, illustrating the influence of meta-covariates on the covariate effects. The GC-content and the length of the genes seem to have an impact mainly on the effect of the number of reads mapped to introns and those with ambiguous mapping.

We focus the remaining part of this section on the results obtained thanks to the innovative treatment of the residual term allowed by \textsc{cosin}. 
The adaptive Gibbs sampler identifies $15$ active factors. 
The rank-one contributions $C_h$ allow one to decompose the underlying signal in rank-one additive matrices which aid interpretation as discussed henceforth.

Figure \ref{fig:cont-covid} displays, as an example, the estimated $C_1$, $C_2$, $C_3$, and $C_4$ matrices. To facilitate interpretation, we re-arranged the column order according to the COVID-19 pathway in explaining similarities among genes.
The first contribution positively impacts the expression of the genes related to the COVID-19 pathway. The second contribution influences the expression of the genes in an opposite way, yet still indicating the importance of the pathway information to decompose the residual term.
Such an evidence of strong association between the COVID-19 pathway and lung cancer cells may suggest a similar inflammation pathway for lung adenocarcinoma and COVID-19. In fact, COVID-19 promotes activation of the NF-$\kappa$B pathway via Ang II type 1 receptor (AT1R), followed by interleukin-6 (IL-6) production \citep{perico2021immunity}. Such activation of the innate immune system, which triggers overproduction of pro-inflammatory cytokines, including IL-6, can result in systemic inflammatory response \citep{perico2021immunity}.
A similar process happens in lung cancer, in which immune response and cytokines play an important role. For instance, overexpression of IL-6 is associated with tumor progression through inhibition of cancer cell apoptosis, stimulation of angiogenesis, and drug resistance \citep{guo2012interleukin}.
A further indication of the similar molecular mechanisms of COVID-19 and cancer is the reporpusing of several cancer treatments as experimental treatments for severe COVID-19 \citep{jafarzadeh2020contribution}. 
The third and fourth contributions are characterized by a completely different pattern. Their effect is no longer homogeneous across the rows, but it reveals the existence of a cell stratification not captured by the mean of the process with respect to the cell line.

\begin{figure}[h!t]
	\centering
	\includegraphics[width=.9\textwidth]{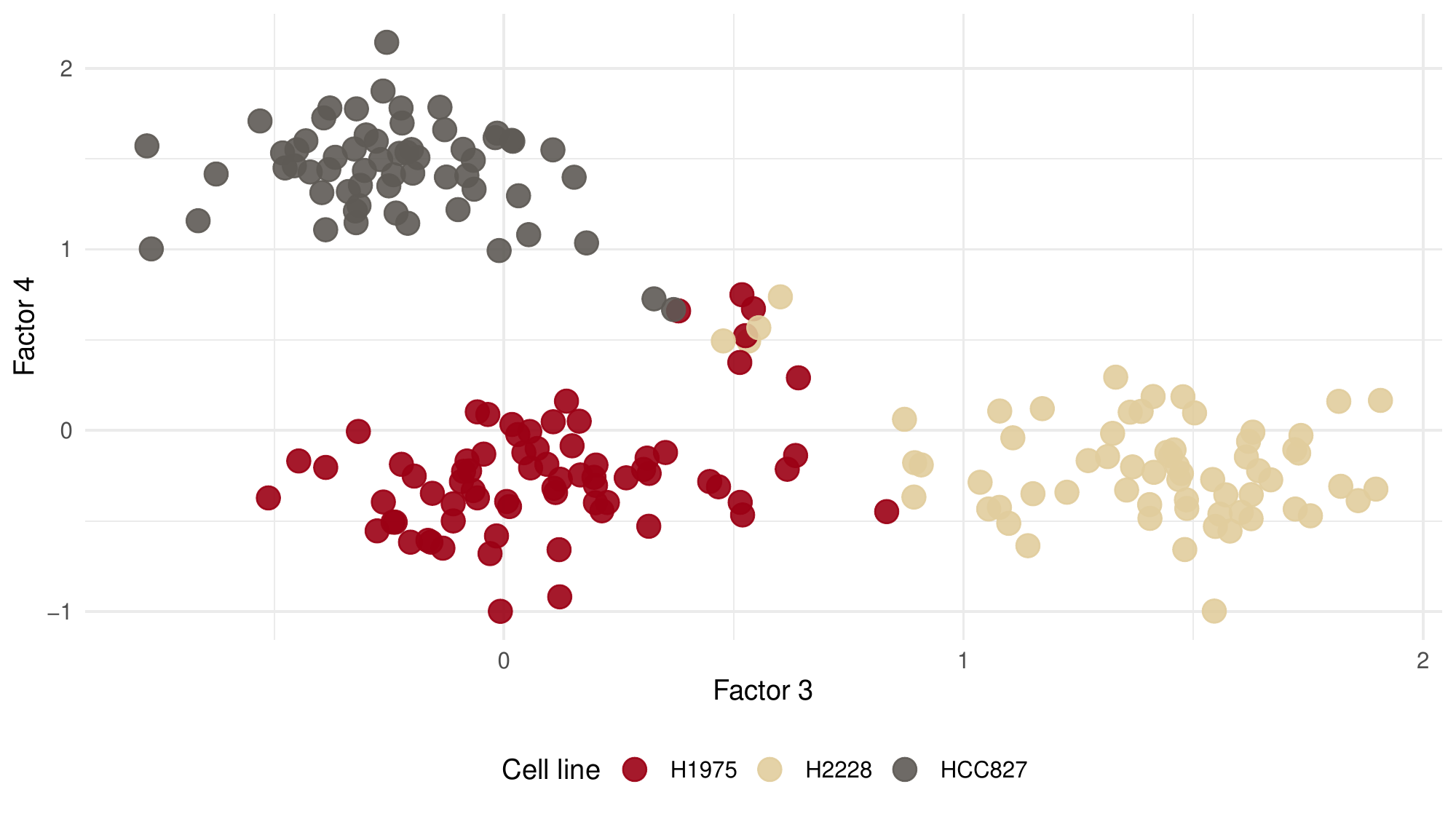}
	\caption{Representation of the $199$ cells plotted on the plane identified by factor scores $\eta_{\cdot 3}$ and $\eta_{\cdot 4}$. Points are coloured according to the cell lines.}
	\label{fig:cell-lines}
\end{figure}

To investigate the ability of the model in recognizing possibly interpretable latent unobserved covariates, in Figure \ref{fig:cell-lines} we plot a representative posterior draw of the cell factor scores $\eta_{\cdot 3}$ and $\eta_{\cdot 4}$, chosen following the recommendations of Section \ref{sec:posteriorcomputation}. Cells are coloured according to the cell lines in order to assess correspondence with the three clusters revealed by our approach. We can appreciate a clear pattern suggesting that the structure imposed on the latent part of the model aid in recognizing and reconstructing possible missing covariates.

Notably, the contributions that explain heterogeneity among cells are generally not characterized by a differentiated impact on the genes with respect to the biological pathways. In other terms, considering the full model specification for the data point $y_{ij}$
\begin{equation*}
\lfloor y_{ij} \rfloor = \exp (z_{ij}) = \exp(x_{i}^\top \beta_j) \, \bigg\{\prod_{h=1}^{15} \exp(C_{hij})\bigg\} \exp(\varepsilon_{ij}),
\end{equation*}
we observe a multiplicative factorization in different contributions of the cell line role and the genes pathway role in characterizing the gene expression. 

Latent factors not discussed here are more difficult to interpret. However, they contribute to explain the residual variance, and may be possibly important in guiding new biological discoveries in future studies.

\begin{figure}[h!t]
	\centerline{
		\includegraphics[width=.9\textwidth]{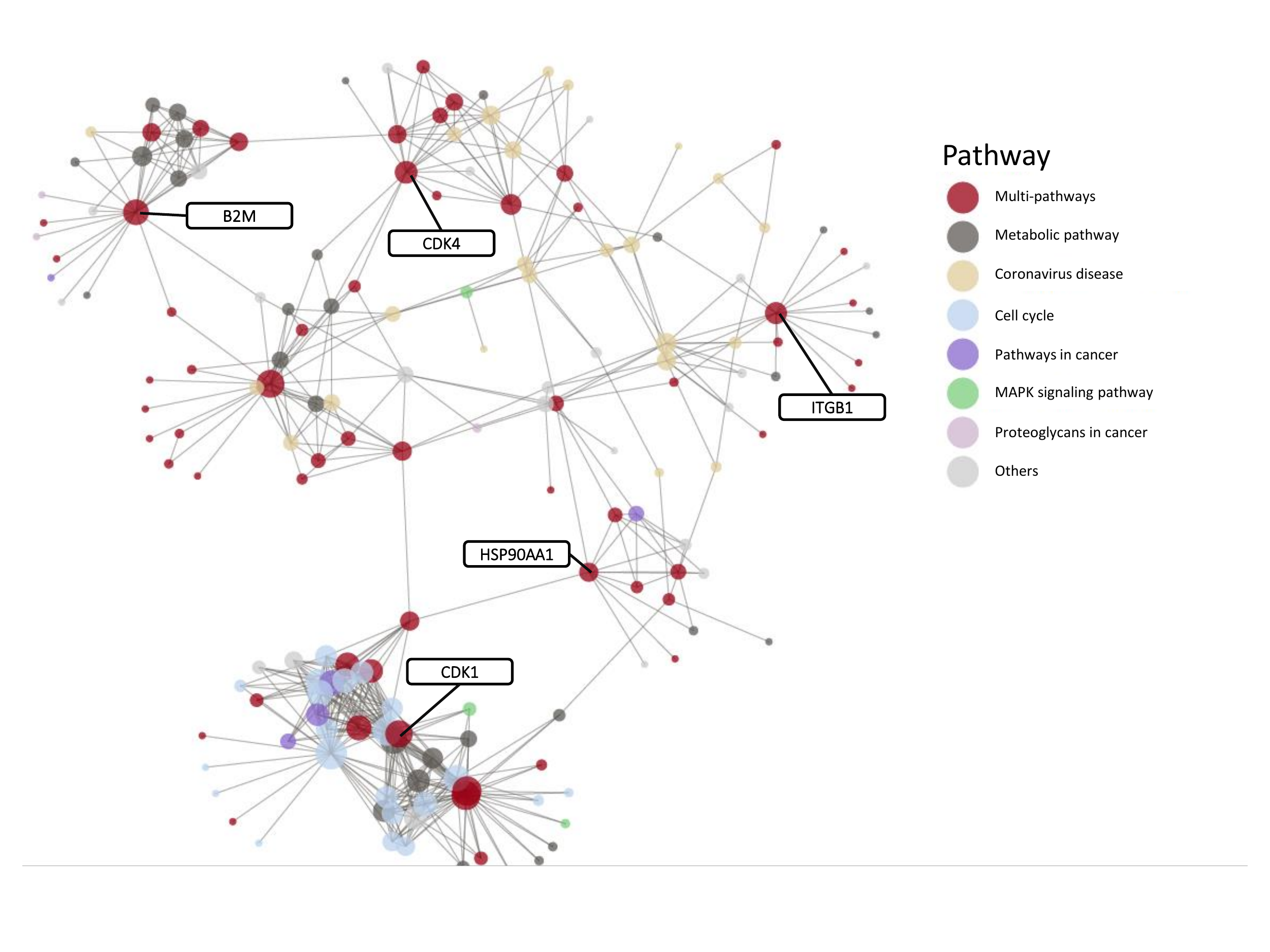}}
	\caption{Graphical representation based on the posterior mean of the inverse of the correlation matrices estimated by the model. Edge thicknesses are proportional to the latent partial correlations between genes. Values below 0.025 are not reported.
		Nodes are positioned using a Fruchterman-Reingold force-direct algorithm.}
	\label{fig:genes-net}
\end{figure}

Additional insights can be gained by exploring the structure of the sparse covariance matrix $\Omega = \Lambda \Lambda^\top + \Sigma$. 
The graph constructed from the posterior mean of its inverse (i.e. the partial correlation matrix), reported in Figure \ref{fig:genes-net}, reveals the presence of gene communities with genes belonging to the same pathway having the tendency of being clustered together. For instance, the cluster observed at the bottom of the graph is mainly constituted by the genes belonging to the cell cycle and metabolic and cancer pathways, while the genes related to the COVID-19 pathway are mainly distributed in the communities at the top right corner of the graph.
This structure is favoured, yet not imposed, by the dependence on the meta-covariates induced by the structured increasing shrinkage specification on the loadings matrix $\Lambda$.
In addition, the graph highlights the genes which stand out as hub nodes, since are correlated with large groups of genes, or link different clusters. Notable hub genes include cell cycle kinases, such as CDK1 and CDK4, essential for G1/S and G2/M phase transitions, and for cell cycle control \citep{nurse1998understanding}. Because of the central role of the cell cycle in cancer progression, these genes have been proposed as therapeutic targets for inhibitors in lung adenocarcinoma treatment \citep{asghar2015history}. 
Other hub nodes that have been associated with cancer in the literature include HSP90AA1 and ITGB1, which belong to pathways in cancers and to the PI3K-AKT signaling pathway, and B2M that is involved in immune resistance promoting cancer transformation of the cells and escape from therapies \citep{pereira2017}.

\section{Simulation study}
\label{s:simulations}
To illustrate the validity and generality of \textsc{cosin}, we assess its performances through a simulation study. As competitor, we use the generalized principal component analysis \citep[or \textsc{glm-pca},][]{townes2019}, representing the current state-of-the-art in dimension reduction for sRNA-seq data. We are interested in evaluating both the predictive capacity of the model and its ability in recognizing and isolating the underlying signal. 
In order to assess the main novel aspects of our approach, which lie in how count data and residual error terms are treated, we consider the zero mean data generating process
\begin{align*}
y_{ij} &= \lfloor \exp(z_{ij}) \rfloor, \,  z_{ij} =  C_{1ij}+ C_{2ij} +C_{3ij} +\varepsilon_{ij},\,
\varepsilon_{ij} \sim N(0,\sigma^2).
\end{align*}
To mimic the situation observed in the application discussed in Section \ref{s:application}, we induce row-wise and column-wise sparsity over the contribution matrices $C_2$ and $C_3$, respectively. In particular, we specify $C_{hij}=\eta_{ih} \lambda_{jh}$ with
\begin{align*}
\eta_{\cdot1}, \eta_{\cdot 3} &\sim N_n(0,I_n), \quad \lambda_{\cdot1}, \lambda_{\cdot2} \sim N_p(0,I_p),\\
\eta_{i2} &\sim N(0,0.05^2), \quad i> n/2,  \quad \eta_{l2}=1, \quad l \leq n/2,\\
\lambda_{j3} &\sim N_p(0,0.05^2), \quad j> p/2,  \quad \lambda_{m3}=1, \quad m \leq p/2.
\end{align*}
Information on sparse columns is provided to the competing models through a single meta-covariate vector $w_B$ containing entries $w_j$ equal to one for $j\leq p/2$ and zero otherwise. The role of such a meta-covariate is analogous to the role of a biological genes-specific meta-covariate in the scRNA-seq data application.
\textsc{cosin} and \textsc{glm-pca} are also estimated ignoring meta-covariate information to assess their robustness to the missing of informative variable traits.
On the other hand, row-sparsity simulates the unobserved cell stratification and then should be entirely inferred via model estimation.

Six different scenarios are obtained varying the data dimensions $(n,p)$ over the set $\{(50,100),$ $ (200,100), (200,1000)\}$ and idiosyncratic variance $\sigma^2$ over $\{0.1^2, 1\}$.
For each scenario, we simulate $50$ synthetic data sets.

To assess goodness-of-fit, we perform an out-of-sample prediction task, randomly removing a sample $\mathcal{S}$ of $25\%$ of entries in $y$. We fit \textsc{cosin}, \textsc{cosin} without meta-covariates, and \textsc{glm-pca} ignoring entries of  $\mathcal{S}$  and compute the mean absolute error of model $m$ defined as 
$$
\text{MAE}_m=\frac{1}{|\mathcal{S}|}\sum_{l \in \mathcal{S}}\left|y_l - \hat{y}^{(m)}_l\right|,
$$
where $\hat{y}^{(m)}_l$ is the value predicted by the model $m$. 

For \textsc{cosin}, we set hyperparameters $\alpha=5$,  $\sigma_\gamma^2=1$, $a_\theta=b_\theta=1$, $a_\sigma=b_\sigma = 1$, and $c_p=0.5$. Posterior distribution is approximated via the Gibbs algorithm reported in  Appendix A in the supporting information. We compute the MAE of \textsc{cosin} averaging over all the MAE calculated on the basis of the parameters sampled at every iteration of the Gibbs.
Unlike our proposal, the competitor model \textsc{glm-pca} is not equipped of a methodology to infer the number of latent components $k$, which should be provided before the estimation. To favour a fair comparison, we estimate the \textsc{glm-pca} models under different values of $k \in  \{2,3,4,5,6,7,8\}$, using as a benchmark the specification characterized by the lowest MAE.

\begin{figure}[h!t]
	\centering
	\includegraphics[width=.99\textwidth]{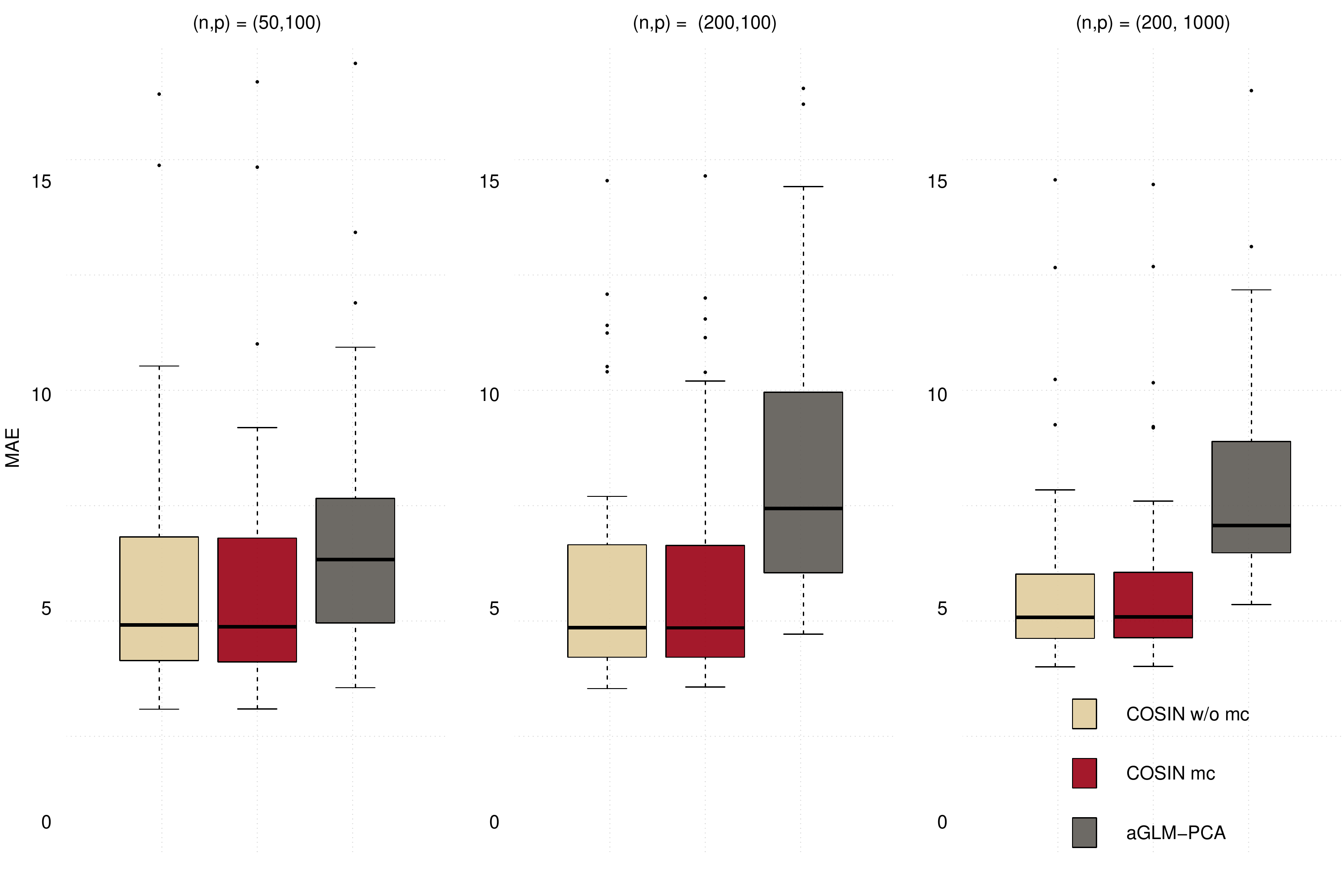}
	\caption{Boxplot of the out-of-sample MAE of the competing models under different values of $(n,p)$ with $\sigma^2=1$. Since the {\sf R} package \texttt{glmpca} does not allow for out-of-sample imputation, we adopt the \textsc{glm-pca} approximation proposed by \citet{townes2019}, i.e. we perform principal component analysis on Pearson residuals of the null Poisson model.}
	\label{fig:sim1}
\end{figure}

Figure \ref{fig:sim1} provides a summary of the results for scenarios with $\sigma^2 =1$. Our proposed approach shows the best performance across all scenarios. Although meta-covariates have a minor impact on model fitting, they play a crucial role in providing valuable and interpretable estimates, as discussed in Section \ref{s:application}. The full results are presented in  Table A4, reported in the Appendix.

Finally, we evaluated  the ability of the different methods in reconstructing the original underlying signal. This assessment is based on the difference between the original rank-one contributions and the contributions estimated by the competing approaches. In particular, Table \ref{tab:RMSE} reports the root mean squared errors on the three contributions under the \textsc{cosin} and \textsc{glm-pca} models fitted with meta-covariates on the full data matrices used for the predictive comparison. The estimated contributions of each model are ordered by minimizing the sum of contributions RMSE, while the number of latent components $k$ in \textsc{glm-pca} models is set equal to the value minimizing MAE.
For \textsc{cosin}, RMSE reported is the average of the RMSE computed at every Gibbs iteration.
Results point out that the latent constructs identified by \textsc{glm-pca} are far from the original contributions generating the data, suggesting that the use of the \textsc{cosin}  can bring some advantages in decoupling the several layers that explain the residual variance in a highly multivariate context.

\begin{table}
	\centering
	\caption{Median of contribution RMSE in 50 replicates, with varying $(n,p,\sigma)$. Interquartile range is reported in parenthesis.  \label{tab:RMSE}}
	\begin{tabular}{lrrrrrr}
		$(n,p,\sigma)$  & \multicolumn{3}{c}{\textsc{cosin}} & \multicolumn{3}{c}{\textsc{glm-pca}}  \\
		&  $C_1$ & $C_2$ & $C_3$ & $C_1$ & $C_2$ & $C_3$ \\
		\hline
		$(50,100,0.10)$ & 0.63 & 0.58 & 0.39 & 1.85 & 1.76 & 0.84 \\
		& (0.25) & (0.28) & (0.26) & (2.13) & (1.25) & (0.18) \\
		$(50,100,1.00)$ &  0.62 & 0.57 & 0.39 & 0.63 & 0.79 & 0.76 \\
		& (0.30) & (0.26) & (0.14) &  (0.17) & (0.16) & (0.13) \\
		$(200,100,0.10)$ & 0.54 & 0.45 & 0.25 &  1.56 & 1.44 & 0.74 \\ 
		&  (0.35) & (0.33) & (0.19) &  (0.58) & (0.78) & (0.03) \\
		$(200,100,1.00)$ &  0.48 & 0.46 & 0.26 & 0.48 & 0.66 & 0.74 \\
		&  (0.41) & (0.32) & (0.05) &  (0.12) & (0.11) & (0.07) \\
		$(200,1000,0.10)$ &  0.87 & 0.47 & 0.63 &  0.99 & 0.86 & 0.72 \\
		&  (0.12) & (0.08) & (0.06) & (0.23) & (0.42) & (0.05) \\
		$(200,1000,1.00)$ & 0.62 & 0.55 & 0.53 &  0.81 & 0.68 & 0.73 \\
		&  (0.12) & (0.07) & (0.04) & (0.61) & (0.12) & (0.05) \\
	\end{tabular}
	\vspace*{6pt}
\end{table}

\section{Discussion}
\label{s:discuss}
In this study, we introduced a novel method called \textsc{cosin} which provides a joint modeling approach for multivariate count data through latent factor models. This approach was specifically motivated by scRNA-Seq applications, which involves complex count data. The empirical performances observed in real and synthetic data sets demonstrate that \textsc{cosin} shows competitive results in terms of model fitting compared to existing methods, indicating its efficacy as a modeling framework.

One key advantage of \textsc{cosin} is that it allows for the modeling of latent sparsity through the use of meta-covariates. This feature is particularly useful in identifying latent contributions that have  biological interpretation. As discussed in Section 3, \textsc{cosin} was successfully applied to a scRNA-seq dataset on lung adenocarcinoma, where it identified specific latent contributions $C_j$ that were associated with different cell types and pathways. These findings highlight the potential utility of \textsc{cosin} for uncovering biologically meaningful patterns in high-dimensional count data.

One important aspect to consider is the role of the intercept in the prior mean of the coefficients $\beta_j$. Indeed, this is a gene-wise intercept that helps the model accounts for the differences in sequencing depth across cells, similarly to what is achieved with the gene-wise intercept in GLM-PCA \citep{townes2019} and with offsets in more traditional frequentist models \citep{robinson2010edger,love2014moderated}.

Clearly, our results suggest that \textsc{cosin} is applicable beyond genomics and can be used in any context dealing with complex high-dimensional counts. This versatility makes \textsc{cosin} a valuable tool for researchers working in diverse fields.


\section*{Acknowledgements}
Davide Risso and Giovanni Toto are supported by the National Cancer Institute of the National Institutes of Health (U24CA180996). Davide Risso is also supported by EU funding within the MUR PNRR ``National Center for HPC, big data and quantum computing'' (Project no. CN00000013 CN1).


%

\begin{thebibliography}{}
	
	\bibitem[\protect\citeauthoryear{Anders and Huber}{Anders and
		Huber}{2010}]{anders2010}
	Anders, S. and Huber, W. (2010).
	\newblock Differential expression analysis for sequence count data.
	\newblock {\em Nature Precedings} {\bf 11,}.
	
	\bibitem[\protect\citeauthoryear{Asghar, Witkiewicz, Turner, and
		Knudsen}{Asghar et~al.}{2015}]{asghar2015history}
	Asghar, U., Witkiewicz, A.~K., Turner, N.~C., and Knudsen, E.~S. (2015).
	\newblock The history and future of targeting cyclin-dependent kinases in
	cancer therapy.
	\newblock {\em Nature Reviews Drug Discovery} {\bf 14,} 130--146.
	
	\bibitem[\protect\citeauthoryear{A{\ss}mann, Boysen-Hogrefe, and
		Pape}{A{\ss}mann et~al.}{2016}]{assmann2016}
	A{\ss}mann, C., Boysen-Hogrefe, J., and Pape, M. (2016).
	\newblock Bayesian analysis of static and dynamic factor models: An ex-post
	approach towards the rotation problem.
	\newblock {\em Journal of Econometrics} {\bf 192,} 190--206.
	
	\bibitem[\protect\citeauthoryear{Becker, Nevins, Chen, Chiu, Horning, Guha,
		Laquindanum, Mills, Chaib, Ladabaum, et~al\mbox{.}}{Becker
		et~al.}{2022}]{becker2022single}
	Becker, W.~R., Nevins, S.~A., Chen, D.~C., Chiu, R., Horning, A.~M., Guha,
	T.~K., Laquindanum, R., Mills, M., Chaib, H., Ladabaum, U., et~al. (2022).
	\newblock Single-cell analyses define a continuum of cell state and composition
	changes in the malignant transformation of polyps to colorectal cancer.
	\newblock {\em Nature Genetics} {\bf 54,} 985--995.
	
	\bibitem[\protect\citeauthoryear{Bhattacharya and Dunson}{Bhattacharya and
		Dunson}{2011}]{bhattacharya2011}
	Bhattacharya, A. and Dunson, D.~B. (2011).
	\newblock Sparse bayesian infinite factor models.
	\newblock {\em Biometrika} {\bf 98,} 291--306.
	
	\bibitem[\protect\citeauthoryear{Canale and Dunson}{Canale and
		Dunson}{2011}]{canale2011}
	Canale, A. and Dunson, D.~B. (2011).
	\newblock Bayesian kernel mixtures for counts.
	\newblock {\em Journal of the American Statistical Association} {\bf 106,}
	1528–--1539.
	
	\bibitem[\protect\citeauthoryear{Guo, Xu, Lu, Duan, and Zhang}{Guo
		et~al.}{2012}]{guo2012interleukin}
	Guo, Y., Xu, F., Lu, T., Duan, Z., and Zhang, Z. (2012).
	\newblock Interleukin-6 signaling pathway in targeted therapy for cancer.
	\newblock {\em Cancer Treatment Reviews} {\bf 38,} 904--910.
	
	\bibitem[\protect\citeauthoryear{Jafarzadeh, Chauhan, Saha, Jafarzadeh, and
		Nemati}{Jafarzadeh et~al.}{2020}]{jafarzadeh2020contribution}
	Jafarzadeh, A., Chauhan, P., Saha, B., Jafarzadeh, S., and Nemati, M. (2020).
	\newblock Contribution of monocytes and macrophages to the local tissue
	inflammation and cytokine storm in covid-19: Lessons from sars and mers, and
	potential therapeutic interventions.
	\newblock {\em Life Sciences} {\bf 257,} 118102.
	
	\bibitem[\protect\citeauthoryear{Jiang, Sun, Song, and Li}{Jiang
		et~al.}{2022}]{jiang2022}
	Jiang, R., Sun, T., Song, D., and Li, J.~J. (2022).
	\newblock Statistics or biology: the zero-inflation controversy about scrna-seq
	data.
	\newblock {\em Genome Biology} {\bf 23,} 1--24.
	
	\bibitem[\protect\citeauthoryear{Khatri, Sirota, and Butte}{Khatri
		et~al.}{2012}]{khatri2012ten}
	Khatri, P., Sirota, M., and Butte, A.~J. (2012).
	\newblock Ten years of pathway analysis: current approaches and outstanding
	challenges.
	\newblock {\em PLoS Computational Biology} {\bf 8,} e1002375.
	
	\bibitem[\protect\citeauthoryear{Kowal and Canale}{Kowal and
		Canale}{2020}]{kowal}
	Kowal, D.~R. and Canale, A. (2020).
	\newblock Simultaneous transformation and rounding (star) models for
	integer-valued data.
	\newblock {\em Electronic Journal of Statistics} {\bf 14,} 1744--1772.
	
	\bibitem[\protect\citeauthoryear{Legramanti, Durante, and Dunson}{Legramanti
		et~al.}{2020}]{legramanti2020}
	Legramanti, S., Durante, D., and Dunson, D.~B. (2020).
	\newblock Bayesian cumulative shrinkage for infinite factorizations.
	\newblock {\em Biometrika} {\bf 107,} 745--752.
	
	\bibitem[\protect\citeauthoryear{Love, Hogenesch, and Irizarry}{Love
		et~al.}{2016}]{love2016modeling}
	Love, M.~I., Hogenesch, J.~B., and Irizarry, R.~A. (2016).
	\newblock Modeling of rna-seq fragment sequence bias reduces systematic errors
	in transcript abundance estimation.
	\newblock {\em Nature Biotechnology} {\bf 34,} 1287--1291.
	
	\bibitem[\protect\citeauthoryear{Love, Huber, and Anders}{Love
		et~al.}{2014}]{love2014moderated}
	Love, M.~I., Huber, W., and Anders, S. (2014).
	\newblock Moderated estimation of fold change and dispersion for rna-seq data
	with deseq2.
	\newblock {\em Genome Biology} {\bf 15,} 1--21.
	
	\bibitem[\protect\citeauthoryear{Marioni, Mason, Mane, Stephens, and
		Gilad}{Marioni et~al.}{2008}]{marioni2008}
	Marioni, J.~C., Mason, C.~E., Mane, S.~M., Stephens, M., and Gilad, Y. (2008).
	\newblock Rna-seq: an assessment of technical reproducibility and comparison
	with gene expression arrays.
	\newblock {\em Genome Research} {\bf 18,} 1509--1517.
	
	\bibitem[\protect\citeauthoryear{McParland, Gormley, McCormick, Clark,
		Kabudula, and Collinson}{McParland et~al.}{2014}]{mcparland2014}
	McParland, D., Gormley, I.~C., McCormick, T.~H., Clark, S.~J., Kabudula, C.~W.,
	and Collinson, M.~A. (2014).
	\newblock Clustering south {African} households based on their asset status
	using latent variable models.
	\newblock {\em The Annals of Applied Statistics} {\bf 8,} 747.
	
	\bibitem[\protect\citeauthoryear{Nurse, Masui, and Hartwell}{Nurse
		et~al.}{1998}]{nurse1998understanding}
	Nurse, P., Masui, Y., and Hartwell, L. (1998).
	\newblock Understanding the cell cycle.
	\newblock {\em Nature Medicine} {\bf 4,} 1103--1106.
	
	\bibitem[\protect\citeauthoryear{O'Hara and Kotze}{O'Hara and
		Kotze}{2010}]{hara2010}
	O'Hara, R. and Kotze, J. (2010).
	\newblock Do not log-transform count data.
	\newblock {\em Nature Precedings} pages 1--1.
	
	\bibitem[\protect\citeauthoryear{Ovaskainen and Abrego}{Ovaskainen and
		Abrego}{2020}]{ovaskainen2020}
	Ovaskainen, O. and Abrego, N. (2020).
	\newblock {\em Joint species distribution modelling: with applications in R}.
	\newblock Cambridge University Press.
	
	\bibitem[\protect\citeauthoryear{Pereira, Gimenez-Xavier, Pros, Pajares, Moro,
		Gomez, Navarro, Condom, Moran, Gomez-Lopez, et~al\mbox{.}}{Pereira
		et~al.}{2017}]{pereira2017}
	Pereira, C., Gimenez-Xavier, P., Pros, E., Pajares, M.~J., Moro, M., Gomez, A.,
	Navarro, A., Condom, E., Moran, S., Gomez-Lopez, G., et~al. (2017).
	\newblock Genomic profiling of patient-derived xenografts for lung cancer
	identifies b2m inactivation impairing immunorecognitionb2m inactivation in
	lung cancer affects immune recognition.
	\newblock {\em Clinical Cancer Research} {\bf 23,} 3203--3213.
	
	\bibitem[\protect\citeauthoryear{Perico, Benigni, Casiraghi, Ng, Renia, and
		Remuzzi}{Perico et~al.}{2021}]{perico2021immunity}
	Perico, L., Benigni, A., Casiraghi, F., Ng, L.~F., Renia, L., and Remuzzi, G.
	(2021).
	\newblock Immunity, endothelial injury and complement-induced coagulopathy in
	covid-19.
	\newblock {\em Nature Reviews Nephrology} {\bf 17,} 46--64.
	
	\bibitem[\protect\citeauthoryear{Polson, Scott, and Windle}{Polson
		et~al.}{2013}]{polson2013}
	Polson, N.~G., Scott, J.~G., and Windle, J. (2013).
	\newblock Bayesian inference for logistic models using p{\'o}lya--gamma latent
	variables.
	\newblock {\em Journal of the American Statistical Association} {\bf 108,}
	1339--1349.
	
	\bibitem[\protect\citeauthoryear{Risso, Schwartz, Sherlock, and Dudoit}{Risso
		et~al.}{2011}]{risso2011gc}
	Risso, D., Schwartz, K., Sherlock, G., and Dudoit, S. (2011).
	\newblock Gc-content normalization for rna-seq data.
	\newblock {\em BMC Bioinformatics} {\bf 12,} 1--17.
	
	\bibitem[\protect\citeauthoryear{Roberts and Rosenthal}{Roberts and
		Rosenthal}{2007}]{roberts2007}
	Roberts, G.~O. and Rosenthal, J.~S. (2007).
	\newblock Coupling and ergodicity of adaptive markov chain monte carlo
	algorithms.
	\newblock {\em Journal of Applied Probability} {\bf 44,} 458--475.
	
	\bibitem[\protect\citeauthoryear{Robinson, McCarthy, and Smyth}{Robinson
		et~al.}{2010}]{robinson2010edger}
	Robinson, M.~D., McCarthy, D.~J., and Smyth, G.~K. (2010).
	\newblock edger: a bioconductor package for differential expression analysis of
	digital gene expression data.
	\newblock {\em Bioinformatics} {\bf 26,} 139--140.
	
	\bibitem[\protect\citeauthoryear{Robinson and Smyth}{Robinson and
		Smyth}{2008}]{robinson2008}
	Robinson, M.~D. and Smyth, G.~K. (2008).
	\newblock Small-sample estimation of negative binomial dispersion, with
	applications to sage data.
	\newblock {\em Biostatistics} {\bf 9,} 321--332.
	
	\bibitem[\protect\citeauthoryear{Roy, Borg, and Dunson}{Roy
		et~al.}{2021}]{roy2019}
	Roy, A., Borg, J.~S., and Dunson, D.~B. (2021).
	\newblock Bayesian time-aligned factor analysis of paired multivariate time
	series.
	\newblock {\em The Journal of Machine Learning Research} {\bf 22,}
	11347--11373.
	
	\bibitem[\protect\citeauthoryear{Schiavon, Canale, and Dunson}{Schiavon
		et~al.}{2022}]{schiavon2022}
	Schiavon, L., Canale, A., and Dunson, D.~B. (2022).
	\newblock Generalized infinite factorization models.
	\newblock {\em Biometrika} {\bf 109,} 817--835.
	
	\bibitem[\protect\citeauthoryear{Tian, Dong, Freytag, L{\^e}~Cao, Su,
		JalalAbadi, Amann-Zalcenstein, Weber, Seidi, Jabbari, et~al\mbox{.}}{Tian
		et~al.}{2019}]{tian2019}
	Tian, L., Dong, X., Freytag, S., L{\^e}~Cao, K.-A., Su, S., JalalAbadi, A.,
	Amann-Zalcenstein, D., Weber, T.~S., Seidi, A., Jabbari, J.~S., et~al.
	(2019).
	\newblock Benchmarking single cell rna-sequencing analysis pipelines using
	mixture control experiments.
	\newblock {\em Nature Methods} {\bf 16,} 479--487.
	
	\bibitem[\protect\citeauthoryear{Tian, Su, Dong, Amann-Zalcenstein, Biben,
		Seidi, Hilton, Naik, and Ritchie}{Tian et~al.}{2018}]{scpipe}
	Tian, L., Su, S., Dong, X., Amann-Zalcenstein, D., Biben, C., Seidi, A.,
	Hilton, D.~J., Naik, S.~H., and Ritchie, M.~E. (2018).
	\newblock {scPipe}: A flexible {R}/{B}ioconductor preprocessing pipeline for
	single-cell {RNA}-sequencing data.
	\newblock {\em {PLOS} {C}omputational {B}iology} {\bf 14,} e1006361.
	
	\bibitem[\protect\citeauthoryear{Townes, Hicks, Aryee, and Irizarry}{Townes
		et~al.}{2019}]{townes2019}
	Townes, F.~W., Hicks, S.~C., Aryee, M.~J., and Irizarry, R.~A. (2019).
	\newblock Feature selection and dimension reduction for single-cell rna-seq
	based on a multinomial model.
	\newblock {\em Genome biology} {\bf 20,} 1--16.
	
	\bibitem[\protect\citeauthoryear{Wagner, Regev, and Yosef}{Wagner
		et~al.}{2016}]{wagner2016revealing}
	Wagner, A., Regev, A., and Yosef, N. (2016).
	\newblock Revealing the vectors of cellular identity with single-cell genomics.
	\newblock {\em Nature Biotechnology} {\bf 34,} 1145--1160.
	
	\bibitem[\protect\citeauthoryear{Warton}{Warton}{2018}]{warton2018}
	Warton, D.~I. (2018).
	\newblock Why you cannot transform your way out of trouble for small counts.
	\newblock {\em Biometrics} {\bf 74,} 362--368.
	
	\bibitem[\protect\citeauthoryear{Wu and Dai}{Wu and Dai}{2017}]{wu2017tumor}
	Wu, T. and Dai, Y. (2017).
	\newblock Tumor microenvironment and therapeutic response.
	\newblock {\em Cancer Letters} {\bf 387,} 61--68.
	
	\bibitem[\protect\citeauthoryear{Xue, Zhao, Aronowitz, Mai, Vides, Qeriqi, Kim,
		Li, de~Stanchina, Mazutis, et~al\mbox{.}}{Xue et~al.}{2020}]{xue2020rapid}
	Xue, J.~Y., Zhao, Y., Aronowitz, J., Mai, T.~T., Vides, A., Qeriqi, B., Kim,
	D., Li, C., de~Stanchina, E., Mazutis, L., et~al. (2020).
	\newblock Rapid non-uniform adaptation to conformation-specific kras (g12c)
	inhibition.
	\newblock {\em Nature} {\bf 577,} 421--425.
	
\end{thebibliography}

\newpage 

\makeatletter
\renewcommand\thefigure
{\@arabic\c@figure}
\def\fps@figure{tbp}
\def\ftype@figure{2}
\def\fstyle@figure{\reset@font\small\rm}
\def\ext@figure{lof}
\def\fnum@figure{{\bf Figure A\thefigure}
}
\makeatother

\makeatletter
\def\fnum@table{\textbf{Table A\thetable}}
\makeatother

\section*{Appendix}

Appendix A includes the Gibbs sampling algorithm referenced in Sections \ref{s:application}--\ref{s:simulations}.
In Appendix B, referenced in Section \ref{s:application},  Tables and a  Figure related to the lung adenocarcinoma scRNA-seq data application are displayed.
 Appendix C, referenced in Section \ref{s:simulations}, reports a Table about the simulation experiments.

An implementation of the Gibbs sampling algorithm is included in the {\sf R} package {\sf cosin} available at \url{https://github.com/giovannitoto/cosin}.
The {\sf R} code developed and used in the paper is available at \url{https://github.com/giovannitoto/cosin_code}. 
\vspace*{-8pt}

\subsection*{Appendix A}

In this appendix we report the Gibbs sampling algorithm for the \textsc{cosin} approach referenced in Sections 3--4 of the paper.
Detailed settings of the algorithm are also included.

Given $k^*$ the number of factors at iteration $t$, the algorithm iterates along the following steps
\begin{enumerate}[ label= \textit{\roman*.}, leftmargin=2\parindent]
	
	\item Update the elements $z_{ij}$ ($i=1,\ldots,n$; $j=1\ldots,p$) sampling independently and in parallel from the truncated normal
	\begin{equation*}
		(z_{ij} \mid -) \sim TN(x_{i}^T \beta_{jh} + \sum_{h=1}^{k}\eta_{ih}\lambda_{jh},  1, l_{ij}, u_{ij}), 
	\end{equation*}
	with lower bound $l_{ij} = \log(y_{ij})$ and upper bound $u_{ij}= \log(y_{ij}+1)$. 
	
	\item Update in parallel the $d$ technical meta-covariate coefficient row vectors $\gamma_{lT}^\top$ ($l=1,\ldots,d$) sampling from conditionally independent posteriors
	\begin{equation*}
		(\gamma_{lT}\mid-) \sim N_{q_T}\big\{
		(I_{q_T} + \sigma_\beta^{-2} w_T^\top w_T)^{-1} \sigma_\beta^{-2}(w_T^\top \beta_l),\,(I_{q_T} + \sigma_\beta^{-2} w_T^\top w_T)^{-1}\big\},
	\end{equation*}
	where $\beta_l^\top$ is the $l$th $p$-variate row vector of the coefficient matrix $\beta$.	
	
	\item Update in parallel the $p$ covariate coefficient vectors $\beta_j\;(j=1, \ldots,p)$ by sampling from the independent full conditional posterior distributions 
	\begin{equation*}
		(\beta_{j} \mid -) \sim N_d \big[ 
		(\sigma_\beta^{-2} I_d + x^\top x)^{-1} \{x^\top (z_{j}-\eta \lambda_{j} )+ \sigma_j^{-2} \Gamma_T w_{jT} \},(\sigma_\beta^{-2} I_d + x^\top x)^{-1}\big],
	\end{equation*}
	where $\eta$ is the $n \times k^*$ matrix of latent scores and $z_{j}$ is the $j$th column vector of $z_j = [z_{1j}, \ldots, z_{nj}]^\top$.
	Then, we set $\epsilon = z - x \beta$.
	
	\item Update, for $i=1, \ldots, n$, the factor $\eta_i$ according to the posterior full conditional 
	\begin{equation*}
		(\eta_{i}\mid-) \sim N_{k^*}\big\{(I_{k^*}+\Lambda_{k^*}^\top \Sigma^{-1} \Lambda_{k^*})^{-1} \Lambda_{k^*}^\top \Sigma^{-1} \epsilon_{i},\, (I_{k^*}+\Lambda_{k^*}^\top \Sigma^{-1} \Lambda_{k^*})^{-1}\big\}.
	\end{equation*} 
	The distribution is conditional to $\Lambda$ and $\epsilon$, so we can update in parallel the $n$ vectors $\eta_i$.
	
	\item Update in parallel the $p$ elements of $\Sigma$, by sampling 
	\begin{equation*}
		(\sigma_{j}^{-2}\mid-) \sim \text{Ga}\left\{a_\sigma +\frac{n}{2},\, b_\sigma +\frac{1}{2} \sum_{i=1}^{n} (\epsilon_{ij}-\lambda_{j}^\top \eta_i)^2 \right \}.
	\end{equation*} 
	
	\item Update the local scale parameters proceeding as follows.
	\begin{enumerate}[label* = \roman*]		
		\item Let $\phi_{jh} =\varphi_{jh} \tilde{\phi}_{jh}$, with $\varphi_{jh}$ and $\tilde{\phi}_{jh}$ independent a priori and distributed as 
		Ber$\{\text{logit}^{-1}(w_{jB}^\top \gamma_{hB})\}$ and Ber$(c_p)$, respectively.
		Update $\varphi_{jh}$, for $j=1,\ldots,p$ and $h=1,\ldots,k^*$, setting $\varphi_{jh}=1$ if $\phi_{jh}=1$ and sampling from the full conditional distribution
		\begin{equation*}
			\text{pr}(\varphi_{jh}=u) \propto 
			\begin{cases}
				1-\text{logit}^{-1}(w_{jB}^\top \gamma_{hB})\qquad \quad \;\;\, \text{for} \;
				u=0, \\
				\text{logit}^{-1}(w_{jB}^\top \gamma_{hB})(1-c_{p})  \qquad \,   \text{for} \;u=1,\\
			\end{cases}
		\end{equation*}		
		if $\phi_{jh}=0$.
		Given $w_{B}\Gamma_{B}$, the elements $\varphi_{jh}$ ($j=1,\ldots,p$, $h=1,\ldots,k^*$) are independently distributed and can be updated in parallel. 
		
		\item Update each column vector $\gamma_{hB}$ of $\Gamma_{B}$ exploiting the P\'{o}lya-Gamma data-augmentation strategy \citep{polson2013}.
		Let $\textnormal{pr}(x)\propto \sum_{n=0}^{\infty} (-1)^n A_n (2\pi x^3)^{-0.5}\exp\{-(2n+b)^2(8x)^{-1}-0.5c^2x\}$ indicate the probability density function of a P\'{o}lya-Gamma distributed random variable $x \sim \text{PG}(b,c)$.
		For each $h=1, \ldots, k^*$, generate $p$ independent random variables $d_{j(h)}$ sampling from the full conditional distribution $(d_{j(h)}\mid -) \sim \text{PG}(1, w_{jB}^\top \gamma_{hB}^{(t-1)})$.
		Let $D_{(h)}$ denote the $p\times p$ diagonal matrix with entries $d_{j(h)}$ ($j=1,\ldots,p$).
		For each $h=1, \ldots, k^*$, update $\gamma_{hB}$ sampling from
		\begin{equation*}
			(\gamma_{hB} \mid -) \sim N_{q_B}\{(w^\top D_{(h)} w + I_{q_B} )^{-1} (w_{B}^\top \kappa_h), \, (w_{B}^\top D_{(h)} w + I_{q_B} )^{-1}\},
		\end{equation*}
		where $\kappa_{h}$ is a $p$-dimensional vector with the $j$-th entry equal to $\varphi_{jh}-0.5$. Given $\varphi_{jh}\,(j=1,\ldots,p;\;h=1,\ldots,k^*)$, we can update in parallel all the vectors $\gamma_{hB}\, (h=1,\ldots,k^*)$.
	\end{enumerate}
	
	\item  Let $\tilde{\lambda}_{jh}$ denote the continuous underlying loadings element such that $\lambda_{jh}=\phi_{jh} \rho_h \tilde{\lambda}_{jh}$ ($h=1,\ldots,k^*$). 
	Then, update the elements $\tilde{\lambda}_{jh}$ by sampling from the independent full conditional posterior distributions of the row vector $\tilde{\lambda}_{j}$  ($j=1,\ldots,p$),
	\begin{equation*}
		(\tilde{\lambda}_{j}\mid-) \sim \mathcal{N}_{k^*}\big\{(D_{j}^{-1}+F_{j} \eta^\top \eta  F_{j})^{-1} F_{j} \eta^\top  \epsilon_{j}, \, (D_{j}^{-1}+F_{j} \eta^\top \eta F_{j})^{-1}\big\},
	\end{equation*} 
	where $D^{-1}_j=\text{diag}(\vartheta_1^{-1},\ldots, \vartheta_{k^*}^{-1})$, $F_j = \text{diag}( \rho_1 \phi_{1j}, \ldots,\rho_{k^*} \phi_{k^*})$,
	and $\epsilon_j$ is the column vector $\epsilon = [\epsilon_{1j}, \ldots, \epsilon_{nj}]^\top$. 
	The distribution is conditional to factor scores $\eta$ and to the scale matrix and vector $\Phi$ and $\theta$, thus we can update in parallel the $p$ vectors $\tilde{\lambda}_j$.
	Finally, set $\lambda_{jh}=\phi_{jh} \rho_h \tilde{\lambda}_{jh}$ for any $j=1,\ldots,p$ and $h=1,\ldots,k^*$.
	
	\item Update the factor-specific scale parameters proceeding as follows.
	\begin{enumerate}[label* = \scshape .\roman*]				
		\item Following \citet{legramanti2020}, define the independent indicators $\xi_h$ ($h=1,\ldots,p$) with prior $\text{pr}(\xi_h=l)= u_l$.
		Update the augmented data $\xi_h$ by sampling from the full conditional distribution
		\begin{equation*}
			\text{pr}(\xi_h=l) \propto 
			\begin{cases}
				u_l \,\textnormal{pr}_{N}\{\text{vec}(\epsilon); \text{vec}(\eta\Lambda) - \text{vec}(\eta_h\lambda_h^\top) , I_{n p}\}
				\qquad \text{for} \quad l=1,\ldots,h \\
				u_l \, \textnormal{pr}_{N}\{\text{vec}(\epsilon); \text{vec}(\eta\Lambda)+ (1-\rho_h^{(t-1)}) \text{vec}(\eta_h \lambda^{*\top}_h) ,I_{n p}\} \qquad  \text{for} \quad l=h+1,\ldots,k^*, 
			\end{cases}
		\end{equation*}
		where we define the row vector $\lambda^{*\top}_h$ such that $\lambda_h^{\top} = \rho_h \lambda^{*\top}_h$.
		Then, $\rho_h=1$ if $\xi_h >h$, else $\rho_h=0$.
		The full conditional distribution of $\xi_h$ depends on the value of $\rho_l\,(l=1,\ldots,k^*)$, implying to immediately set $\lambda_h^{\top} = \rho_h \lambda^{*\top}_h$ and to update $\rho_h$ sequentially with respect to the index $h=1,\ldots,k^*$.
		
		\item Update the parameter $\vartheta_h$ ($h=1,\ldots,k^*$) by sampling $\vartheta_h^{-1}$ from the full conditional distribution $\text{Ga}(a_\theta+0.5p, b_\theta+0.5\sum_{j=1}^{p} \tilde{\lambda}_{jh}^{2})$.
		
		\item For $l=1,\ldots,k^*-1$, sample $v_l$ from 
		\begin{equation*}
			(v_l\mid-) \sim \text{Be}\big\{1+ \sum_{h=1}^{k^*} \mathbbm{1}{(\xi_h=l)}, \alpha + \mathbbm{1}{(\xi_h>l)} \big\},
		\end{equation*}
		while set $v_{k^*} = 1$.
		Since, given $u$, the distribution are conditionally independents, we can update the $k^*$ elements of the vector $v$ in parallel. 
		Finally, update $u_l=v_l \prod_{m=1}^{l-1} (1-v_m)$, for $l=1,\ldots,k^*$. 
	\end{enumerate} 
	
	\item Update the local scale by sampling from the full conditional distributions of $\phi_{jh}$.
	If $h \in \{1,\ldots,k^*: \rho_h=0\}$, then sample from the Bernoulli defined as
	\begin{equation*}
		\text{pr}(\phi_{jh}=\xi) = \begin{cases}
			1-\text{logit}^{-1}(w_{jB}^\top \gamma_{hB})\,c_{p}\, 
			\qquad  \; \text{for} \; \xi=0\\
			\text{logit}^{-1}(w_{jB}^\top \gamma_{hB})\,c_{p} \qquad \qquad \, \text{for} \;  \xi=1.
		\end{cases} 
	\end{equation*}
	Given the linear predictor matrix $w_{B} \Gamma_{B}$, we can update in parallel $\phi_{jh}$ for $j=1,\ldots,p$ and $h \in \{1,\ldots,k^*: \rho_h=0\}$.
	If $h \notin \{1,\ldots,k^*: \rho_h=0\}$, sample from
	\begin{equation*}
		\text{pr}(\phi_{jh}=\xi) \propto \begin{cases}
			\{1-\text{logit}^{-1}(w_{jB}^\top \gamma_{hB})\,c_{p}\}\,\textnormal{pr}_{N}\left(\epsilon_j; \eta \lambda_j - \phi_{jh}^{(t-1)} \tilde{\lambda}_{jh} \eta_{h} , I_{n}\right)
			\, \quad  \; \text{for} \; \xi=0\\
			\text{logit}^{-1}(w_{jB}^\top \gamma_{hB})\,c_{p} \, \textnormal{pr}_{N}\left\{\epsilon_j;  \eta\lambda_j +\left(1- \phi_{jh}^{(t-1)}\right) \tilde{\lambda}_{jh} \eta_{h}  ,I_{n}\right\}  \quad \text{for} \;  \xi=1.
		\end{cases} 
	\end{equation*}
	where $\textnormal{pr}_{N}(x;\mu, I_n)$ is the multivariate density function of the $n$-variate Gaussian distribution with mean $\mu$, variance equal to the identiy matrix, and evaluated at $x$. We use $\phi_{jh}^{(t-1)}$ to denote the parameter $\phi_{jh}$ sampled at the previous iteration of the Gibbs and $\eta_{h}$ indicating the $h$th factor score vector.
	The distribution of each $\phi_{jh}$ depends on $w_{B} \Gamma_{B}$ and on the elements $\phi_{lj}$ ($l=1,\ldots,h$) via $\lambda_{j}$. Therefore, the update is sequential with respect to the index $h$ and requires to set $\lambda_{jh}=\phi_{jh} \tilde{\lambda}_{jh}$ after having sampled $\phi_{jh}$ for any $h \notin \{1,\ldots,k^*: \rho_h=0\}$. On the other hand, we can update in parallel with respect to the index $j=1,\ldots, p$.
	
\end{enumerate}

The results reported in lung adenocarcina scRNA-seq data application in Section 3 and in the simulations studies in Section 4 are obtained running the algorithms for 20000 iterations discarding the first 5000 iterations. 
Then, we thin the Markov chain, saving every 2-th sample.
We adapt the number of active factors at iteration $t$ with probability $\textnormal{pr}(t) = \exp(-1 -5 \, 10^{-4} t)$ after at least 100 iterations.
In simulation studies, we consider a single vector $w_B$ of biological meta-covariates as described in the main paper. 

\clearpage
\subsection*{Appendix B}

This appendix includes additional tables and figures for lung adenocarcinoma scRNA-seq data application referenced in Section 3 of the paper.

\begin{table}[ht]
	\def\~{\hphantom{0}}
	\centering
	\begin{threeparttable}
		\caption{List of biological pathways considered as meta-covariates} \label{tab:pathways}
		\begin{tabular}{lr}
			Pathway & Number of genes  \\
			\hline
			Metabolic pathways & $313$\\
			Pathways of neurodegeneration multiple diseases & $153$\\
			Amyotrophic lateral sclerosis & $121$ \\
			Alzheimer disease & $116$\\
			Pathways in cancer & $109$ \\
			Parkinson disease & $108$  \\
			COVID 19 pathway & $103$\\
			Salmonella infection & $100$\\
			Shigellosis & $92$\\
			PI3K-AKT signaling pathway & $79$ \\
			Human papillomavirus infection & $77$\\
			Huntington disease & $73$\\
			Regulation of actin cytoskeleton & $71$\\
			Cell cycle & $69$\\
			Focal adhesion & $65$\\
			Tight junction & $64$\\
			MAPK signaling pathway & $63$\\
			Human cytomegalovirus infection & $63$\\
			Human immunodeficiency virus 1 infection & $62$\\
			Proteoglycans in cancer & $62$\\
		\end{tabular}
	\end{threeparttable}
\end{table}

\begin{table}[ht]
	\def\~{\hphantom{0}}
	\centering
	\begin{threeparttable}
		\caption{Summary results of the posterior mean of $\beta$ matrix and number of genes characterized by $0.9$-level credible intervals not including zero.} \label{tab:cov-coeff}
		\begin{tabular}{lrrrrrr}
			Covariate & Min. & Q$_0.25$  & Q$_0.5$ & Q$_0.75$&  Max. & $0.9$-level genes \\
			\hline
			Intercept & $-0.504$ & $0.089$ & $ 0.164 $ & $ 0.279 $ & $ 0.579 $ & $108$\\
			Unaligned & $-0.371 $ & $-0.062 $ & $ 0.004 $ & $ 0.067 $ & $ 0.326 $ & $0$ \\
			Aligned unmapped & $ -0.627 $ & $ -0.067 $ & $ 0.01 $ & $ 0.071 $ & $ 0.413 $ & $17$\\
			Mapped to exon &$ -0.218 $ & $ -0.013 $ & $ 0.037 $ & $ 0.088 $ & $ 0.338 $ & $0$ \\
			Mapped to intron & $ -0.296 $ & $ -0.075 $ & $ -0.009 $ & $ 0.055 $ & $ 0.378 $ & $24$ \\
			Ambiguous mapping & $ -0.379 $ & $ -0.051 $ & $ 0.018 $ & $ 0.086 $ & $ 0.369 $ & $0$ \\
			Mapped to MT & $ -0.231 $ & $ -0.045 $ & $ -0.001 $ & $ 0.044 $ & $ 0.292 $ & $23$\\
			Number of genes  & $ -0.338 $ & $ 0.050 $ & $ 0.117 $ & $ 0.189 $ & $ 0.517 $ & $13$\\
			Total count & $ -0.162 $ & $ 0.012 $ & $ 0.051 $ & $ 0.093 $ & $ 0.261 $ & $0$\\
		\end{tabular}
	\end{threeparttable}
\end{table}

\begin{figure}[ht]
	\centerline{
		\includegraphics[width =0.85\columnwidth]{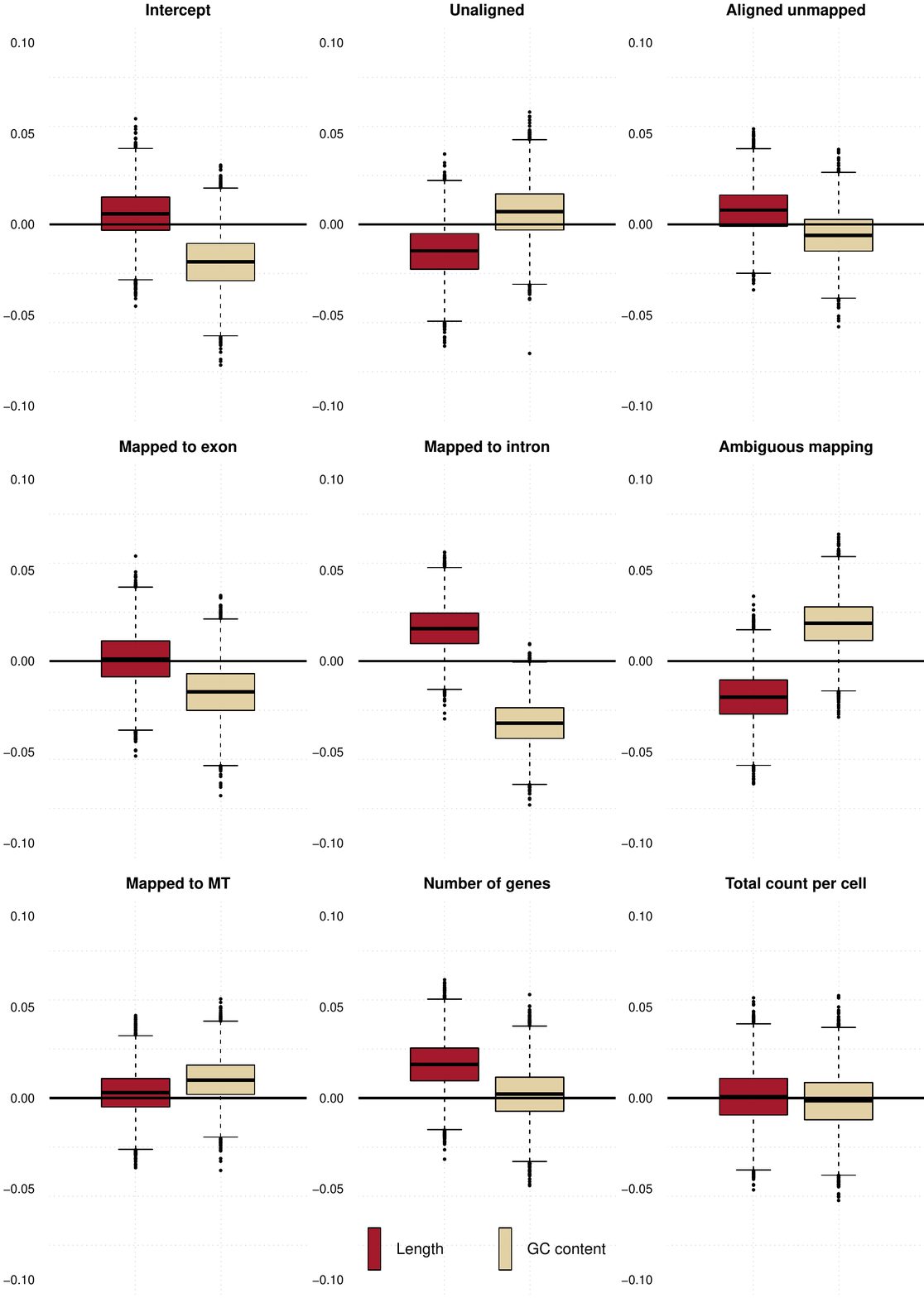}}
	\caption{Posterior distributions of the parameter $\Gamma_T$ entries. Red and yellow boxplots indicate the influence of the genes length and the genes GC content, respectively, on the covariate effects.}
	\label{fig:cov-metacov}
\end{figure}

\clearpage
\subsection*{Appendix C}

In this appendix we report the full table referenced in Section 4 of the paper reporting the predictive performances of the two competing model.

\begin{table}[ht]
	\centering
	\def\~{\hphantom{0}}
	\begin{threeparttable}
		\caption{Median of MAE in 50 replicates, with  varying $(n,p,\sigma)$. Interquartile range is reported in parenthesis.  \label{tab:MAE}}
		\begin{tabular}{rccc}
			$(n,p,\sigma)$ & \textsc{cosin} w/o metacov. & \textsc{cosin}  metacov. & \textsc{glm-pca}$^*$ w/o metacov. \\
			\hline
			$(50,100,0.10)$ & 0.7464 & 0.7544 & 1.4730 \\
			& (0.5243) & (0.5076) & (1.4275) \\
			$(50,100,1.00)$ & 4.6107 & 4.5624 & 6.1408 \\
			& (2.8316) & (2.8434) & (2.8640) \\
			$(200,100,0.10)$ & 0.5055 & 0.5071 & 1.3518 \\
			& (0.2754) & (0.2776) & (1.1313) \\
			$(200,100,1.00)$ & 4.5456 & 4.5388 & 7.3334 \\ 
			& (2.6220) & (2.6028) & (4.1877) \\
			$(200,1000,0.10)$ & 0.4196 & 0.4170 & 1.4331 \\
			& (0.1119) & (0.1197) & (0.7778) \\
			$(200,1000,1.00)$ & 4.7887 & 4.7921 & 6.9329 \\
			& (1.4916) & (1.5049) & (2.4602) \\
		\end{tabular}
		\begin{tablenotes}
			\small
			\item $^*$ Since the {\sf R} package \texttt{glmpca} does not allow for out-of-sample imputation, we adopt the \textsc{glm-pca} approximation proposed by \citet{townes2019}, i.e. we perform principal component analysis on Pearson residuals of the null Poisson model. 
		\end{tablenotes}
	\end{threeparttable}
\end{table}

\end{document}